# Enabling Prescription-based Health Apps

**Venet Osmani, Stefano Forti, Oscar Mayora**
Fondazione Bruno Kessler
TrentinoSalute 4.0 Competence Center
Trento, Italy
{vosmani, forti, omayora} @ fbk.eu

**Diego Conforti**
Autonomous Province of Trento
TrentinoSalute 4.0 Competence Center
Trento, Italy
diego.conforti@provincia.tn.it

## ABSTRACT
We describe an innovative framework for prescription of personalised health apps by integrating Personal Health Records (PHR) with disease-specific mobile applications for managing medical conditions and the communication with clinical professionals. The prescribed apps record multiple variables including medical history enriched with innovative features such as integration with medical monitoring devices and wellbeing trackers to provide patients and clinicians with a personalised support on disease management. Our framework is based on an existing PHR ecosystem called TreC, uniquely positioned between healthcare provider and the patients, which is being used by over 70.000 patients in Trentino region in Northern Italy. We also describe three important aspects of health app prescription and how medical information is automatically encoded through the TreC framework and is prescribed as a personalised app, ready to be installed in the patients' smartphone.

## INTRODUCTION
Smartphones and their apps have permeated many aspects of our everyday life. There appears to be an app for almost anything, with varying degree of functionality. In this context, health is no exception. According to IMS Institute for Healthcare Informatics in 2015 there were over 165,000 health apps in Apple and Google app stores. This number represents double of the number of health apps recorded in 2013. This rise in health apps use can be attributed to interest in taking control of one's health and having an increased autonomy in management of their disease. Complementary to this, there is a clear trend towards personalised treatment models, dubbed under the umbrella term "*precision medicine*" [1]. The premise of precision medicine is that successful disease treatment outcomes are dependent on incorporating individual phenotypes into treatment models. This realisation has given rise to personalised treatments, where genotype, as well as environmental and behavioural factors become an integral part of the course of treatment. Along this line, apps are being increasingly used to monitor multiple factors related to the wellbeing and course of treatment. Thus, integration of health apps in the care process will also require personalisation, in order to best support this process.

However, the current personalisation model of health apps resembles that of traditional medicine where the majority of health apps are designed for the majority of consumers, while the degree of personalisation is rudimentary. This was especially evident in a global survey of 1,130 patients and caregivers carried out in the context of get-ehealth EU project [2] where for apps across 5 health categories, namely cancer, diabetes, pain, mental health, wellness, the importance of personalisation [3] emerged as a common challenge. Clearly, personalisation of apps is an important factor, and in our view personalisation should be guided by the existing diagnoses as well as medical advice. In this manner personalisation evolves towards prescription-based apps, akin to prescription of drugs, where medical aspects are the principal focus in this process.

In this paper, we provide an overview of personalisation of medical apps, describe our view towards medical app prescription and describe our framework which enables medical app prescription.

## BEYOND APP PERSONALISATION
Prescribing mobile apps by providers to help keep track of patients' symptoms and provide real-time advice on treatment and medication can reduce costs and medical errors, and also improve the patient experience [4]. In our view, personalisation is an integral part of app prescription. However, app prescription goes beyond solely personalisation, since there are three important aspects to be considered. These aspects have come about as a result of our studies in using the platform to prescribe apps, primarily for patients with Type I diabetes.

*Discretion* is the first aspect we describe. It should be in the discretion of the healthcare provider to customise particular aspects of the app in accordance to characteristics of the condition and the needs of the patient. This is because a system to support remote monitoring of chronically ill patients is not equally applicable to every patient. Each chronic condition requires monitoring of particular aspects and a targeted intervention. A classic example includes changes in glucose levels and recommended plan of action. Furthermore, the healthcare provider would modify app prescription according to patient characteristics, affecting compliance. For example an elderly patient might feel more comfortable and be more adherent to traditional methods of monitoring, such as maintaining a written log to track disease aspects. Placing app prescription at the discretion of healthcare provider gives them the means to prescribe the right app to the right patient and also customise these apps according to patient and disease characteristics.

*Personalisation* is an indispensable aspect to app prescribing. Patients not only require individualised care and assistance but also have different psycho-social abilities. It is therefore necessary to create an app which takes the specific needs of each patient into consideration in order to activate the appropriate functionality.

For example, an alarm for a child with type 1 diabetes would be activated in a different manner with respect to the alarm for a pregnant woman, where glucose targets are different as well as the ability to react to alarms. In the same manner, carbohydrate related functionality is more appropriate for the patients that have the ability to count carbohydrates. For other patients it might be sufficient to have a calendar that tracks glucose levels.

*Temporality* is another essential aspect; remote monitoring or intervention through an app should be defined for a specific period of time, according to the condition of the patient, akin to a prescribed drug. For example, remote monitoring can be activated in the period after the initial diagnosis when patients need to be followed more closely. Continuous monitoring, as is the current practice, may provide data of less interest and increase processing and cognitive load. Provider access to data from the app can be discontinued, and later reactivated for a specific period of time as needed, for example in response to a change of therapy.

During our pilot studies emerged the need to build a technology platform which can support apps offered to patients with chronic diseases, taking into account the requirements set out above. The app would be made available to the patient at the discretion of the healthcare provider, who would customise the app functionality according to patient's needs. The app would be run for a defined period as deemed necessary. The three dimensions outlined above are similar to the process of prescribing a medication, thus app prescription could become part of the medical legislation.

While the above discussion is based on our experience in designing and running a number of pilot trials, we also review the related work below.

**RELATED WORK**

A literature review [5] has shown that there is very little work carried out towards medical app prescription that fulfils the criteria outlined in Section 0. This is despite the fact that one of the best indicators of the importance of app prescription is the willingness of patients to accept them. This is illustrated from the survey of 2000 patients by Digitas Health, where it was found that over 90% of patients were willing to accept an offer of a mobile app from their provider rather than a written paper prescription [6]. In this line, Mount Sinai hospital has created the RxUniverse[1] platform, which is curated list of existing apps on the App Stores that healthcare providers can use to prescribe to patients. BlueStar Diabetes app developed by WellDoc was approved by FDA for patients with type 2 diabetes. The app provides patients with personalised feedback and instructions regarding medications, lifestyle and self-management techniques. The BlueStar app helps patients keep track of blood glucose levels, medications, diet, and exercise while sharing their progress with their health care provider. In developing this app, Quinn et al. have carried out a study of 26 patients where control patients were asked to submit their blood glucose levels every 2 weeks to their providers. Intervention arm patients, on the other hand, received a Bluetooth-enabled OneTouch ultra BG meter, with testing supplies, and a cell phone equipped with WellDoc, Inc, proprietary DiabetesManager software that had logs sent electronically to WellDoc provider at least every 4 weeks. Patients in the intervention group experienced a greater decrease in haemoglobin A1C values (2.03%) compared with the control group (0.68%) at 3 months [7].

In a similar fashion, but on a different context (that of low socioeconomic strata - SES), Wayne et al. [8] conducted a pilot study (n=19) on a smartphone-assisted intervention intended to improve behavioural management of type 2 diabetes in an ethnically diverse, lower SES population within an urban community health setting. The health coach interventions were personalised based on eating, physical activity patterns, and overall health goals. These interventions were created by a student trained in behaviour change techniques. For the 6-months study, the authors reported that the participants showed a mean reduction of 0.43% (SD 0.63) (P=.04) of HbA1c, demonstrating the potential clinical relevance of the intervention.

A study by Mayo Clinic found 20 percent of the patients who attended cardiac rehab and used the app were readmitted to the hospital or visited the emergency department within 90 days, compared with 60 percent of those in the control group [9]. In this study, 44 patients at Mayo Clinic who were hospitalized following a heart attack and stent placement were divided into two groups: 25 received cardiac rehab and the online/smartphone-based program; the 19 in the control group received only cardiac rehab.

Another work on personalised intervention was carried out by Nes et al. [10]. The distinctive aspect of this work was that a therapist formulated personalised feedback, rather than automatically as it is the case with many other apps. The feedback was based on participants registering their eating behaviour, medication taking, physical activities and emotions three times daily using the mobile device. They also registered their fasting blood glucose level in the morning diary.

However, as noted in [5] despite the thousands of health and fitness apps now available for download and the emerging interest in using them for improving health behaviours, very few have been tested in intervention settings. The types of published, peer-reviewed app studies that were available for review were predominantly small sample pilot or feasibility.

Considering that the majority of symptoms of mental disorders are reflected in the behaviour of patients, wearable and mobile sensing technologies can monitor mental disorders. In this respect, a number of apps have been used in the treatment of depression. For example, work by Burns et al. [11] developed the Mobilyze! App which was used in a pilot test (n=8), alongside a website and EMA for adults from the general population. The app used context-awareness to personalise ecological momentary interventions in order to support adherence to therapeutic activities. The authors reported a significant reduction in depression (Mini-International Neuropsychiatric Interview [MINI]: Z=2.15, beta [week]=-.65, P=.03), as well as depression and anxiety symptoms at posttest (Patient Health Questionnaire [PHQ-9]: d=1.95, P<.001; Quick Inventory of Depression Symptoms-Clinician Rated: d=2.28, P<.001; Generalized Anxiety Disorder-7 item scale: d=1.37, P<.001).

A randomised controlled trial (RCT) was carried out by Reid et al. [12] and Kauer et al. [13] in using mobile phones in the early stages of adolescent depression (n=114). The intervention group received self-monitoring of mood, stress and daily activities, while control group received self-monitoring of daily activity only through the app developed on the mobile phone. The authors report no personalisation capability of their app and no significant differences were found at post-test and follow-up on outcomes of depression, anxiety, and stress among adolescents from general practice compared to control group. However, authors conclude that that self-monitoring increases emotional self-awareness, which in turn decreases depressive symptoms for young people with mild or more depressive symptoms.

Watts et al. [14] carried out a RCT (n=35) comparing delivery of Cognitive Behaviour Therapy (CBT) through a mobile phone app, versus a traditional computer in patients with major depression. The authors report no personalisation of their CBT program, however based on their findings, the authors conclude that delivering a CBT program using a mobile application, can result in clinically significant improvements in outcomes for patients with depression. This is because both groups (computer and mobile app) were associated with statistically significant benefits at post-test and this result held also at 3 months follow up, where the depression reduction seen for both groups remained significant.

---

[1] http://www.rxuniverse.com/

## APP-PRESCRIPTION PLATFORM

Our work in creating a framework that allows prescription of apps, has begun with the TreC platform (https://trec.trentinosalute.net/). TreC, which stands for Cartella Clinica del Cittadino (Citizens' Clinical Records) is a platform that enables citizens to access, manage and share information about their health and wellbeing.

TreC is a platform that has been integrated into the health services provided by the Province of Trento. Thus, it is a mature platform that manages the health records of over 70,000 citizens. In addition, TreC is extensible in that it allows development and testing of new services connected to the platform. Thus far the platform has been used to support patients in two distinct areas of healthcare: i) oncology: development of a safe therapy system for the safe delivery of intravenous chemotherapy and a home monitoring system for monitoring and managing toxicity and improving adherence in patients receiving oral anticancer therapies at home [15] and ii) diabetes: a system consisting of a mobile diary and a web dashboard allowing patients to record disease-related information. This information can then be visualised by diabetologists through a dedicated dashboard while also allowing computation of insulin bolus and provide a rule-based alarm functionality [16]. A number of other studies are also running in the platform (data not yet published).

As can be seen from the above description, TreC is a mature and well tested platform that is in a unique position to support medical app prescription, since it manages the health records of the region's citizens. TreC is characterised by an important conceptual feature, whereby, rather than developing a health information hub, TreC has been designed as a "system of systems" with an extensible architecture, where sub-systems can be developed to provide additional and specific functionality.

Using TreC to prescribe apps entails a number of steps that are outlined in Figure 1, where the app is designed, through a drag and drop interface, to contain specific functionality and it is activated and installed in the patient's device. This personalised app then is run over a defined period, where appropriate reminders are sent and data from remote monitoring is gathered.

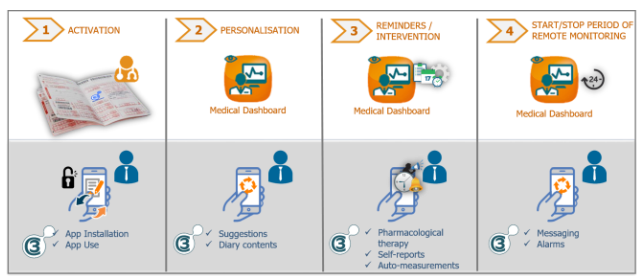

**Figure 1 App-prescription process**

## APP PRESCRIPTION SCENARIO

As we noted in the introduction, in order for the apps to become prescription-based, we need to go beyond the current model of app personalisation, which takes into account only user-supplied data. To address this, our platform is being used by both healthcare providers and patients and has the capability to incorporate medical and lifestyle recommendations. In this respect we have envisioned a scenario, where by a citizen with a chronic disease, such as diabetes, is prescribed an app that will affect their behaviour change.

Maria has been living with type 2 diabetes for the past number of years. She has been struggling with controlling her intake of simple carbohydrates. In order to address this issue, Maria has tried various apps available in the App stores related to diabetes management. However, she found that beyond simple messages of limiting carbohydrate intake and increasing physical activity, the apps did not provide the necessary functionality to manage her condition, nor did they take into account various aspects of her existing medical conditions. In addition to diabetes, Maria suffers from a congenital heart disease, which limits her ability to perform physical activity. Unfortunately, in the apps she downloaded there is no simple way to include all the information regarding her co-morbidities, as listed in her personal health record.

Having been left unsatisfied with the offerings from app stores, Maria visits her primary care doctor and describes her struggle with controlling intake of simple carbohydrates and the limitations of the apps she has tried for the management of Type 2 diabetes. She also tells her doctor that sometimes she forgets to take her morning medications because she wakes up at different times throughout the week.

Together with Maria, the doctor configures the TreC-Diabetes app that can access Maria's health information from her PHR and can incorporate a series of personalised variables for better managing her disease such as Maria's food intolerances, specific physical activity and nutrition targets and the desired frequency of glucose monitoring with her portable monitoring device. In addition, the doctor configures specific messages and reminders to be provided to Maria and sets the date for next visit, all from a single interface.

Through the use of TreC-Diabetes app and TreC framework, the primary care doctor is able to encode all the information told by the patient, while also looking at her medical history for other conditions that the patient may have. This is done through an intuitive menu, where different triggers, conditions and limits can be specified. At the end of the consultation, Maria receives a notification on her smartphone, telling her that the prescribed TreC-Diabetes app is available and ready to be installed. Once Maria installs her pre-configured app she can use it to log her food intake. Her blood glucose levels are automatically transmitted from her glucometer, which has been securely paired with the app. When the app detects that carbohydrate intake is getting near the recommended target, the app notifies Maria to limit further intake of carbohydrates and also recommends physical exercise. The app monitors in the background the amount of physical activity she has performed, using data from her fitness tracking device. Once the safe limits of physical activity have been achieved, as encoded by her primary physician based on her congenital heart disease, the app sends a notification to Maria, recommending her to gradually reduce her exercise intensity. All the data gathered from the app is periodically sent to the TreC platform, where it becomes part of her personal health records, accessible to her primary-care provider. The next morning the app reminds Maria to take her diabetes medication and she begins the day with a feeling of being in control of her condition chronic conditions, with reduced subjective burden of disease.

## CONCLUSION

Smartphones apps have great potential to play an important role in implementation of precision medicine, specifically personalised treatments that consider specific patient aspects. However, as things currently stand, the personalisation of apps is very rudimentary, where only simplistic information is being considered, such as physical characteristics of the patient. Co-existing medical conditions or medical history is typically not taken into consideration.

In order to address these issues we have described our app-prescription platform based on TreC which allows creation and integration of apps on prescription basis. TreC can achieve this because it is uniquely positioned between the healthcare providers and patients. Therefore, it can make use of medical information, such as existing conditions or risk factors in order to design and build apps that take these aspects into consideration. This functionality of prescription-based apps using TreC is already being tested in a pilot study.

## ACKNOWLEDGEMENTS

We would like to acknowledge the contribution of the following people: Stefano Cavallari, Luca Vettoretto, Barbara Purin, Marco Dianti, Flavio Berloffa, Claudio Eccher and Enrico Piras. In addition, we thank Noga Minsky for her valuable feedback on improving this paper.

Ideas presented in this paper served as an input for SmartSDK project (id: 723174), funded by European Commission's Horizon 2020 Programme.